\documentstyle[aaspp4,psfig,12pt]{article}

\def\gtsim{\lower.5ex\hbox{$\; \buildrel > \over \sim \;$}}
\def\ltsim{\lower.5ex\hbox{$\; \buildrel < \over \sim \;$}}
\def\esca{erg s$^{-1}$ cm$^{-2}$ arcmin$^{-2}$}

\begin{document}

\title{Evidence for X-ray emission from a large-scale filament of galaxies?}

\author{Caleb Scharf, Megan Donahue \altaffilmark{1}, G. Mark Voit}
\affil{Space Telescope Science Institute, 3700 San Martin Drive,
Baltimore MD 21218, USA}
%\email{scharf@stsci.edu, donahue@stsci.edu, voit@stsci.edu}

\author{Piero Rosati}
\affil{ESO-European Southern Observatory, D-85748 Garching bei
M\"unchen,  Germany.}
%\email{prosati@eso.org}

\author{Marc Postman}
\affil{Space Telescope Science Institute, 3700 San Martin Drive,
Baltimore MD 21218, USA}
%\email{postman@stsci.edu}

\altaffiltext{1}{Visiting Astronomer, Kitt Peak
National Observatory, National Optical Astronomy Observatories, which is
operated by the Association of Universities for Reasearch in Astronomy, Inc.
(AURA) under cooperative agreement with the National Sciences Foundation.}

\begin{abstract}

Cosmological simulations predict that a large fraction of the baryonic
mass of the Universe exists as $10^{5-7}$ K diffuse, X-ray emitting gas,
tracing low density filament and sheet-like structures exterior to massive
clusters of galaxies. If present, this gas helps reconcile the current
shortfall in observed baryon counts relative to the predictions of the
standard Big Bang model. We present here the discovery and analysis of a
$5 \sigma$ significance half-degree filamentary structure, present in
both I-band galaxy surface density and unresolved X-ray emission in a deep
ROSAT PSPC field. The estimated diffuse X-ray emission component of this
structure has a surface brightness of $\simeq 1.6 \times 10^{-16}$ \esca
(0.5- 2 keV), comparable to the predictions for inter-cluster gas and may
represent a direct detection of this currently unconfirmed baryonic
component. 

\end{abstract}


\section{Introduction}

Low density, large-scale, filamentary and sheet-like structures are seen
in the space distribution of galaxies and are expected in many
models of structure formation (e.g. Cold Dark Matter). Recent simulations
(\cite{cen93,sca93,bry94,cen99}) suggest that low density gas exists in
such structures and can be shock-heated to $10^{5-7}$K without violating
microwave background spectral distortion constraints (\cite{wri94,cen93}).
Furthermore, in order for the observed baryon `budget' (\cite{fuk98}) to
match the detailed predictions of Big Bang nucleosynthesis we might expect
at least $\sim$50\% of present day baryons to be in the form of these warm and
hot plasmas. 

Attempts to detect X-ray emission from filaments that might connect clusters
(\cite{bri95}) have yielded an upper limit to the emission of $< 4\times
10^{-16}$ \esca (at $2-\sigma$ significance). Studies of a likely supercluster
sheet at $z\simeq 0.25$ has yielded evidence for sub-1 keV X-ray gas in fairly
localized, diffuse structure (\cite{wan97}), however the X-ray surface
brightness is $>10$ times higher than that expected from model estimates of the
integrated surface brightness due to warm inter-cluster gas
(\cite{cen93,bry94,cen95,cen99}). These fall in the range of $\sim 1.5 \times
10^{-16}$ \esca, where the major contribution comes from structure at $z\sim
0.2$. Kull \& B\"ohringer (1999) detect evidence for extended emission between
a cluster pair in the Shapley Supercluster, however the emission is $\sim
2.5\times$ brighter than the Briel \& Henry (1995) upper limit and may be due
to cluster merger/interaction rather than genuine intercluster gas.

In order to investigate spatial correlations between galaxies and X-ray
emitting intergalactic gas, we have undertaken a complete optical survey
of the inner regions of 22 deep fields from the ROSAT PSPC archive (the
ROSAT Optical X-ray (ROX) survey, Donahue {\it et al}, in preparation).
Galaxy counts to a completeness limit of $I$-band $m_{I}=23$ have been
obtained for the central $30 \times 30$ arcmin region of each field. 

In addition to seeking distant galaxy clusters via coincidences of galaxy space
overdensities and X-ray emission, this dataset is well suited to investigating
the angular cross-correlation $w(\theta)$ of the unresolved X-ray background
with the distribution of optical galaxies. $w(\theta)$ is a measure of the mean
fractional excess X-ray intensity relative to the mean background at an angle
($\theta$)  from a given galaxy.  Extensive information about the composition
of the X-ray background and the clustering properties of X-ray luminous sources
can then be obtained, see e.g. Refregier {\it et al} (1997), Almaini {\it et
al} (1997). We evaluate $w(\theta)$ independently for each field using a finite
cell estimator (\cite{ref97}, Eqns 2,6).

Seven fields in the ROX survey exhibit positive plateaus in $w(\theta)$ at
angular scales $\ga 2$ arcmin, distinctly different from the expectations
of discrete X-ray sources (\cite{ref97}, Scharf et al, in
preparation).  In one of these fields (labelled CL1603, equatorial
coordinates 16 04 28, +43 13 12 (J2000), ROSAT exposure time 29 ksec) we
have observed a highly extended optical/X-ray structure.  While correlated
optical and X-ray structure is seen in other fields, none exhibit the
apparently contiguous extent of this $\sim 30$ arcmin feature.

\section{Data analysis}

 We have corrected all ROSAT data for telescope vignetting and jitter using
standard exposure maps (\cite{sno92}, and the maximum correction within the
fields is $\sim 15$\%. The 0.5-2 keV counts are used (we have attempted to
perform a similar analysis using the 0.1-0.5 keV band, but the higher
background noise restricts us to measuring uninteresting upper limits), thereby
reducing the background contribution from the Galaxy, minimizing the size of
the point-spread function, and maximizing sensitivity to thermal bressstrahlung
emission from warm and hot gas. 

Before computing $w(\theta)$ for each survey field, or performing the
analyses presented here, we remove all identifiable discrete X-ray sources
and spurious optical features.  We clip, or mask, the X-ray data using a
wavelet-based source detection algorithm (\cite{ros95}) that finds all
sources (extended or point-like) down to a signal-to-noise of $\sim 4$.
The mean surface-brightness is then determined in circular annuli about
each source, and the X-ray source photons are masked to a radius defined
by a surface brightness limit reflecting the background level.  The mean
value of this limit is $\sim 7 \times 10^{-15}$ erg s$^{-1}$ cm$^{-2}$
arcmin$^{-2}$ (0.5-2 keV) over all 22 fields. For the CL1603 field it is
$\simeq 1.9\times 10^{-15}$ erg s$^{-1}$ cm$^{-2}$ arcmin$^{-2}$.  Our
final measurements are not sensitive to the precise specification of this
limit.  Bright stars and scattered light in our optical fields are removed
by clipping out tainted rectangular areas.  The optical and X-ray masks
are then combined and the total mask is applied to {\em both} optical and
X-ray datasets. There are no obvious correlations between the spatial
masking and the apparent filamentary X-ray structure seen in the CL1603
field, as illustrated in Figure 1.

\begin{figure}[htb]
\protect\centerline{\psfig{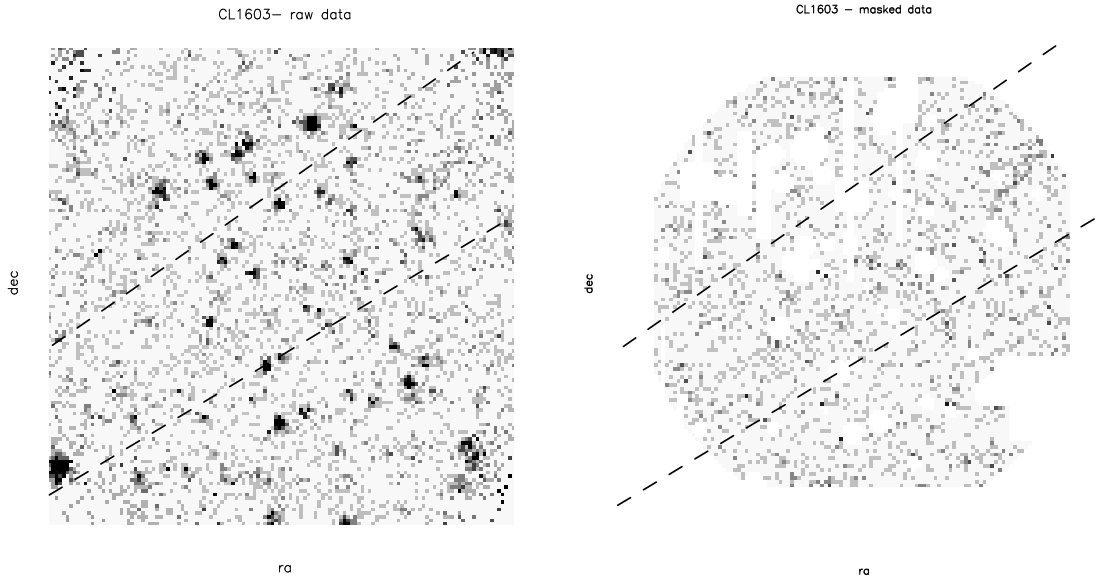}}
\caption{Raw ((a):left panel) and masked ((b): right panel) exposure
corrected, binned (15 arcsec pixels), X-ray total count data for the central
30$\times$30 arcmin region of field CL1603. The lightest greyscale level
(background) corresponds to zero counts, the 2nd lightest to 1 photon/pixel.
(b) illustrates the regions removed by X-ray source excision and those due to
bright stars in the optical data. Dashed lines in (a) \& (b) indicate the
region used for flux estimates of the filamentary structure. The excess flux
associated with this feature can be seen in (b) as a low-level modulation of
the photon distribution.} 
\end{figure}

In order to assess the significance and flux of the  morphologically complex
structure seen in field CL1603 we define its boundary using an aperture
encircling those regions with a galaxy density $> 6 \,  {\rm gal \,
arcmin^{-2}}$ at $I < 22.5$, as shown in Fig. 2.  The structure's flux was
measured only within the part of this aperture which falls between the dashed
lines, and the background  X-ray level was estimated using all data exterior to
the dashed lines. 

\begin{figure}[htb]
\protect\centerline{\psfig{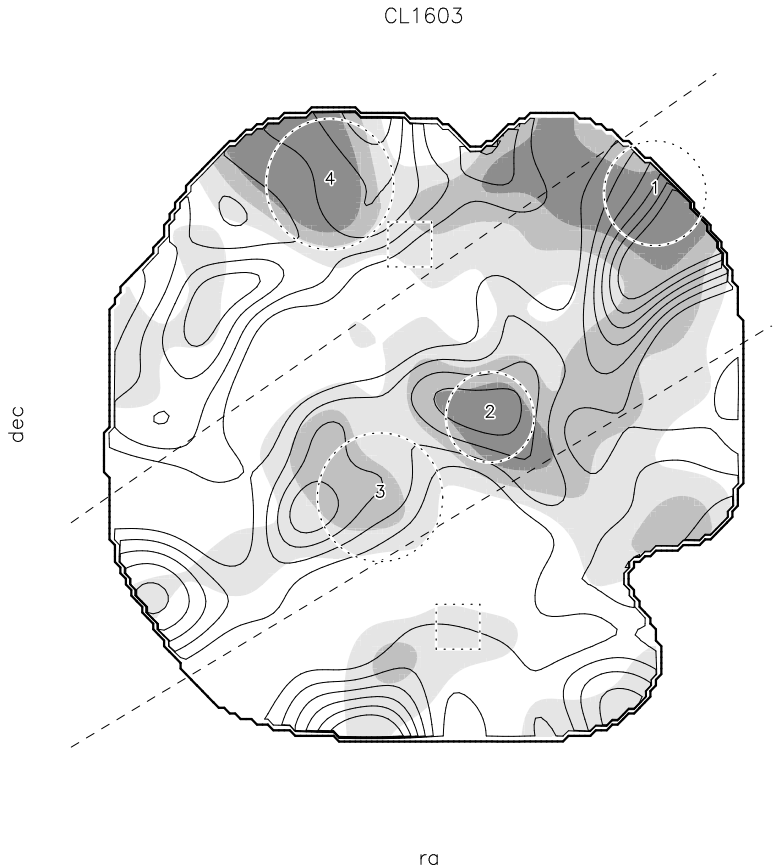}} 
\caption{The smoothed galaxy surface density at $I<22.5$ in the field is
plotted as a 3-level greyscale (lightest level corresponds to $>6$ gal
arcmin$^{-2}$, middle : $>7$ gal arcmin$^{-2}$, darkest: $>8$ gal
arcmin$^{-2}$, highest peaks contain $\sim 10$ gal arcmin$^{-2}$). Smoothed
X-ray surface brightness contours peaked at $2.5\times  10^{-4}$ ct s$^{-1}$
arcmin$^{-2}$ and decreasing in units of  $2\times 10^{-5}$ are overlaid.  Full
masking has been applied to both optical and X-ray data. For both datasets the
smoothing kernel is a gaussian with half-width of $\sigma=1.875$ arcmin, or 7.5
binning units.   To correct for incompleteness the smoothed data is weighted
at  every point by the inverse effective area of the portion of the kernel
containing unmasked data.  Points where this effective area drops below
 $< 30$\% have been discarded,  producing an oddly-shaped outer
boundary. Dashed diagonal lines are as in Figure 1. Numbered circles indicate
the location and extent of  cluster candidates detected with an optical matched
filter  algorithm (note that in the area of no. 4 the X-ray emission and
optical overdensity are not substantially correlated). Squares mark the
location of two previously  identified (\cite{gun86,oke98}) $z\simeq 0.9$
optical cluster candidates.  } 
\end{figure}

The background-subtracted flux of this structure is then $3.1 \times
10^{-3}$ ct s$^{-1}$ (90 counts total), corresponding to $3.6 \times 10^{-14}$
erg s$^{-1}$ cm$^{-2}$ (0.5-2 keV) in an area of 176 arcmin$^2$. The mean
excess surface brightness is thus $2\times 10^{-16}$ erg s$^{-1}$
cm$^{-2}$ arcmin$^{-2}$ Given the background of $1.6 \times 10^{-4}$ ct
s$^{-1}$ arcmin$^{-2}$ (825 counts total in 176 arcmin$^2$), the apparent
filament has signal-to-noise $\sim 3$. 

\section{Statistical significance of faint structure}

An alternative way to assess this structure's significance is to compare the
CL1603 field with the others in the ROX survey, by computing the excess X-ray
flux within similar optically-defined apertures.  This technique takes into
account fluctuations due to all sources, including the `cosmic' variance from
field to field and correlations of optical galaxy counts with pointlike X-ray
sources too faint to be resolved.  We define a flux contrast for each field,
$\Delta X/\bar{X}$, which is the fractional excess of X-ray counts relative to
the expected background within regions containing $> 6 \, {\rm gal \,
arcmin^{-2}}$. This threshold corresponds to a galaxy density somewhat lower
than the mean galaxy density at $I<22.5$ of $7.28$ gal arcmin$^{-2}$, as
determined from the DEEP sky survey (\cite{pos98}).  Our results are relatively
insensitive to this choice but limits outside of the range $5-8$ gal
arcmin$^{-2}$ encompass too little background/source area to compute a sensible
contrast in the fields. The mean flux contrast per field is estimated
together with its dispersion, which we assume to be normally distributed for
this simple analysis.

First, we test the `null' hypothesis that the CL1603 structure is a
random coincidence by evaluating the mean $\Delta X/\bar{X}$ over all
non-matched optical-X-ray combinations of the 22 fields. This null
hypothesis for the CL1603 field is rejected at the $\sim 4\sigma$ level
(Table 1). 

X-ray emission in the Universe is known to be positively correlated with
the large-scale distribution of galaxies (\cite{ref97,alm97}), but we do
not yet know the relative contributions of discrete X-ray sources
(clustered like galaxies) and diffuse, extended sources (gas in groups,
clusters, and larger-scale structures). Following previous work
(\cite{ref97}) we assume that the 15 ROX survey fields {\em without}
excess plateaus in $w(\theta)$ are more representative of cases where the
positively correlated X-ray emission is entirely due to discrete sources
which are themselves galaxies, or at least clustered like the observed
galaxies.  We should therefore expect a systematic (positive) difference
between $\left< \Delta X/\bar{X}\right >$ for the 7 `excess' fields and
that for the 15 other fields. Furthermore, this difference should indicate
the typical fraction of genuinely diffuse X-ray emission which is
positively correlated with the galaxy distribution. 

Table 1 summarizes our results.  As expected, $\Delta
X/\bar{X}$ is systematically lower in the 15 fields with 
no evidence of extended emission from the measurement of $w(\theta)$.
Applying the statistical properties of all 21 other fields to the
CL1603 field, we find $\Delta X/\bar{X}$ to be excessive there
at the $2-3\sigma$ level.  However, these fields could themselves 
contain diffuse filamentary emission.  If instead we establish a
statistical baseline using the 15 fields with no large-scale excess,
the significance of the CL1603 structure rises to $4-5\sigma$.
Taking the results for the 15-field mean (averaging rows 5 and 6 of 
Table 1) as the best estimate of the non-diffuse flux contrast and 
subtracting it from the flux constrast measured for CL1603 we 
estimate that $\sim 77$\% of the flux excess in this field may be 
due to unresolved {\em diffuse} emission. Applying this fraction to our 
previous, direct, estimate of the structure's surface brightness, we
estimate that the diffuse X-ray component has a surface brightness 
of $\sim 1.6 \times 10^{-16}$ erg s$^{-1}$ cm$^{-2}$ (0.5-2 keV).

\begin{table}[htb] \begin{center} \begin{tabular}{lccc} 

Dataset & $\left< \Delta X/\bar{X}\right >$ & Dispersion ($\sigma$) & CL1603
relative significance \\
    & & & \\
CL1603  & 0.256 & - & - \\
Randomized & 0.002 & 0.066 & +3.9$\sigma$ \\    
21 Fields & 0.081 & 0.071 & +2.5$\sigma$ \\
20 Fields & 0.086 & 0.069 & +2.5$\sigma$ \\
(minus 4C23.37) &  &  & \\
15 Fields & 0.058 & 0.045 & +4.4$\sigma$ \\
14 Fields & 0.063 & 0.040 & +4.8$\sigma$ \\  
(minus 4C23.37) &  &  & \\

\end{tabular} \end{center} 
\tablecomments{Table 1: Summary of mean flux contrast ($\left<
\Delta X/\bar{X}\right >$) measurements. For, respectively, the field CL1603,
462 non-repeating randomized optical-X-ray data pairings of all 22 fields, all
21  ROX fields excluding CL1603, all 15 fields not exhibiting large scale
excess in $w(\theta)$ and both datasets not including the field 4C23.37, the
only field to exhibit a negative value of $\Delta X/\bar{X}$. The ($> 6$ gal
arcmin$^{-2}$) optical apertures range in size between 10\% and 75\% of the
total field areas.}

\end{table}

\subsection{Cluster and group flux contributions}

Some of the X-ray emission in the CL1603 field could potentially be coming from
faint, unmasked X-ray clusters.  In order to gauge their effect, we have
applied an optical cluster finding algorithm (\cite{pos96})  to the CL1603
field (Fig. 2 shows detections). None of the candidates are associated with
resolved, extended X-ray sources with fluxes $\ga 1\times 10^{-14}$ erg
s$^{-1}$ cm$^{-2}$, although 2, 3 and 4 are marginally coincident with sources
identified as point-like.  However, they appear to be correlated with both the
galaxy distribution and enhancements in the unresolved X-ray emission. We have
measured the flux in circular apertures defined by the extents of the optical
candidates (Fig. 2) and have estimated the background using all data exterior
to these regions. The signal-to-noise for each candidate is $\la 2$, and the
$2\sigma$ fluxes are $\la 10^{-14}$ erg s$^{-1}$ cm$^{-2}$.  Consequently,
no significant X-ray emission was detected from these individual objects. We
note that excluding the flux within the apertures around objects 1, 2, and 3
reduces the total flux from the putative filament to $2.2\times 10^{-14}$ erg
s$^{-1}$ cm$^{-2}$, however the mean surface brightness is only reduced to $1.6
\times 10^{-16}$ erg s$^{-1}$ cm$^{-2}$ arcmin$^{-2}$. Within the Poisson
uncertainties this is therefore unchanged.

We do not know the distance to the structure in the CL1603 field, but we make
the following observations.  At $18<I<21$ the galaxy density contrast is low
compared to the overall ($I<22.5$) density field, but it appears to trace the
structure similarly.  At $21<I<22$ the density contrast within the structure is
significantly higher and the central regions (near cluster candidates 2 and 3)
are the highest peaks in the field.  For $21 < I < 22$, the typical redshift
would be $z = 0.5 \pm 0.1$ based on the spectrum emitted by a passively
evolving elliptical galaxy (\cite{pos98}). At $22<I<23$ the overall density
fluctuation constrast is reduced, but again, the central regions and the region
close to candidate 1 show the largest enhancements.  Estimated
cluster-candidate redshifts from the cluster-finding algorithm (\cite{pos96}) 
are as follows: (1) $z=0.6$, (2) $z=0.5$, (3) $z=0.3$, (4) $z=0.3$.  Given the
expected success rate of $\sim 70$\% for correctly identifying real clusters
with this method (\cite{pos96}), it seems reasonable that the structure most
likely lies at $z\ga 0.3$ and has a physical extent $\ga 12$
h$_{50}^{-1}$ Mpc (for $h_{50}=H_0/50$km s$^{-1}$ Mpc$^{-1}$ and $q_0 = 0$). 
Our X-ray wavelet detection limits for extended sources ($\sim 1\times
10^{-14}$ erg s$^{-1}$ cm$^{-2}$) implies that the optical cluster candidates
could have $L_x\la 1-4 \times 10^{43}$h$_{50}^{2}$ erg s$^{-1}$ ($q_0=0$).
This further suggests that the observed galaxy enhancement, while possibly
containing very X-ray poor systems, is a genuinely low density, extended
structure.

The original targets of the ROSAT observation, two previously identified
high-redshift ($z\sim0.9$) cluster candidates (\cite{gun86,cas94,oke98}) are
labelled in Fig2. The southern cluster is an unresolved X-ray source
(\cite{cas94})  but not detected optically ($<3\sigma$). The northern cluster
is obscured optically by a nearby bright star and also was not detected in
X-rays. Neither exhibits any strong correlation with the structure.

\section{Discussion}

We have detected an apparent filamentary-like structure at $\sim 3\sigma$
significance in X-ray flux and $\sim 3-5\sigma$ significance as a joint
X-ray/optical overdensity. If real, the structure is likely to be at $z\ga
0.3$ and $\ga 12$h$_{50}^{-1}$Mpc in size. Although optical cluster
candidates are detected in the region, if any are indeed X-ray luminous
then they must be faint, supporting the notion of an extended, low density
system.  Our estimate of {\em diffuse} large-scale emission ($1.6\times
10^{-16}$ erg s$^{-1}$ cm$^{-2}$ arcmin$^{-2}$) is rather lower than
previous constraints on large-scale filamentary X-ray emission (\S 1).

Because estimates of the diffuse X-ray background intensity (\cite{pen99}) are
several times higher than that due to a single filament (\cite{cen93,cen99}),
we estimate that several such filamentary structures could typically be
superposed along any given line of sight, confusing their detection. Given the 
low measured background surface brightness in the CL1603 data (\S 2) we
therefore  suspect that this field may in fact be filament poor, allowing a
single structure to dominate. 

The interpretation of this structure as being predominantly diffuse gas is
clearly not yet secure, albeit theoretically plausible.  Obtaining
photometric and spectroscopic redshift information in this (and
surrounding) fields will be of enormous help in testing the reality of the
observed structures. The CL1603 field is clearly an excellent target for
future X-ray missions such as XMM, where a factor of more than 10 increase
in sensitivity over ROSAT should allow structures such as these to be
studied in detail.

\acknowledgements This research has made use of data obtained
through the High Energy Astrophysics Science Archive Research Center (HEASARC),
provided by the NASA Goddard Space Flight Center. Partial support for this
research  was through NASA LTSA grant NAGW-4502. MD acknowledges the support of
the  STScI Director's Discretionary Research Fund. The authors acknowledge
the planning and data analysis of the project by other members of the 
ROX team, specifically Mark E. Dickinson, Paul Lee, Jennifer Mack, and 
John T. Stocke.

\end{document}